\documentclass[twocolumn,aps,amsmath,nofootinbib,pre,floatfix,superscriptaddress]{revtex4-2}
\usepackage[utf8]{inputenc}
\usepackage{lineno}
\modulolinenumbers[5]
\usepackage{float}
\usepackage{comment}
\usepackage{graphicx}
\usepackage[export]{adjustbox}
\usepackage{wrapfig}
\usepackage{lipsum}  
\usepackage{amsmath}
\usepackage{amssymb}
\usepackage{color}
\usepackage{mdframed}
\usepackage[normalem]{ulem}
\usepackage[thinc]{esdiff}
\usepackage[colorlinks=true, allcolors=blue]{hyperref} 
\usepackage[usenames,dvipsnames,table]{xcolor} 
\usepackage{todonotes}

\usepackage{natbib}

\begin{document}

\title{Seasonal social dilemmas}
\author{Lucas S. Flores }
\affiliation{Instituto de F\'\i sica, Universidade Federal do Rio Grande do Sul, CP 15051, CEP 91501-970 Porto Alegre - RS, Brazil}
\author{Amanda de Azevedo-Lopes}
\affiliation{Department of Theoretical Biology, Max Planck Institute for Evolutionary Biology, Plön 24306, Germany}
\author{Chadi M. Saad-Roy}
\affiliation{Miller Institute for Basic Research in Science, University of California, Berkeley, CA 94720 and Department of Integrative Biology, University of California, Berkeley, CA 94720, USA}
\author{Arne Traulsen}
\affiliation{Department of Theoretical Biology, Max Planck Institute for Evolutionary Biology, Plön 24306, Germany}

\begin{abstract}
Social dilemmas where the good of a group is at odds with individual interests are usually considered as static -- the dilemma does not change over time. In the COVID-19 pandemic, social dilemmas occurred in the mitigation of epidemic spread: Should I reduce my contacts or wear a mask to protect others? In the context of respiratory diseases, which are predominantly spreading during the winter months, some of these situations re-occur seasonally. We couple a game theoretical model, where individuals can adjust their behavior, to an epidemiological model with seasonal forcing. We find that social dilemmas can occur annually and that behavioral reactions to them can either decrease or increase the peaks of infections in a population. Our work has not only implications for seasonal infectious diseases, but also more generally for oscillatory social dilemmas: A complex interdependence between behavior and external dynamics emerges. To be effective and to exploit behavioral dynamics, intervention measures to mitigate re-occuring social dilemmas have to be timed carefully.  
\end{abstract}

\maketitle

\section{Introduction}

While COVID-19 continues to exert a significant burden across the world, 
SARS-CoV-2 has now established itself as an additional circulating pathogen in 
human populations \cite{koelle:science:2022}, joining many other endemic respiratory viruses, 
such as seasonal coronaviruses and influenza. 
For many of these pathogens, 
a hallmark of their epidemiology is that they exhibit seasonal variations in transmission rates \cite{martinez:PLosPath:2018}. 
The drivers of seasonality in infectious disease dynamics range from external factors, 
such as humidity and temperature \cite{shaman:PNAS:2009}, to social factors, 
such as school terms or the timing of large social events  \cite{grassly:PRSB:2006,fares:IJPM:2013}. 
The effect of seasonal drivers on epidemiological dynamics has been examined for a range of specific infectious diseases, ranging from airborne respiratory
infections such as influenza, COVID-19, measles, and chickenpox, to vector-borne diseases,
such as malaria and plague \cite{martinez:PLosPath:2018, faruque:PNAS:2005b, fares:JID:2011, saad-roy:science:2020, earn:science:2000, baker:science:2020}.
Earn et al.\ \cite{earn:science:2000} showed that an epidemic model with seasonal variations in transmission could capture 
key epidemiological transitions in the dynamics of measles transmission. 

In parallel, a key driver of infectious disease dynamics is individual decision-making.
For example, if infection levels or risks from infections are high, individuals may choose to adhere to an 
nonpharmaceutical interventions (NPIs), such as masking or social distancing,
or pharmaceutical interventions, such as vaccination or therapeutics. 
The decision to follow such interventions may then itself change transmission dynamics, 
and affect the circulating number of infections. 
While traditional epidemiological modeling did not explicitly take into account individual reaction to disease 
(e.g. \cite{kermack:PRSA:1927,acedo:NArwa:2010,traulsen:NAL:2022}), 
several studies have examined these effects in more detail
\cite{tanaka:TPB:2002,bauch:PRSB:2005,funk:JRSI:2010,reluga:PLOSCB:2010,glaubitz:PRSA:2020,tanimoto:book:2021,arthur:PLOSCB:2021,arthur:PLOSCB:2023,weitz:PNAS:2020,morsky:PNAS:2023,saad-roy:PNAS:2023, glaubitz:Heliyon:2023}.

Because of the impact of individual behavior on trajectories of the COVID-19 pandemic, 
there has been an increased interest in coupling epidemiological and behavioral models,
and in understanding the potential feedback between these processes \cite{bergstrom:PNAS:2023}. 
Evolutionary game theory provides a mathematical framework to model how individuals adjust their behavior in response to the state of their population.
The decision to comply with a given intervention may lead to social dilemmas, which arise when the individual interests conflict with the collective interests.
One example of such a situation is the prisoner's dilemma (PD) game, where individuals can choose between cooperation and defection.
In the PD game, the best strategy for a player is to defect regardless of the other player's choice. 
This results in a suboptimal scenario where both players defect whereas mutual cooperation would bring a higher gain for both players. 
This scenario is the most common and challenging to study how cooperation can emerge between selfish actors.
Another example is the Snowdrift game (SD), which depicts competition for shared resources and escalation of conflict \cite{maynard-smith:Nature:1973,gore:Nature:2009,hauert:Nature:2004}.
In the SD game, players can maximize their gain by choosing the opposite strategy from the other player -- even if it would be beneficial for the collective if both cooperated. 
The dynamics of such a population are most commonly described by the replicator dynamics \cite{taylor:MB:1978,zeeman:LN:1980}. 
Recent work has shown that this approach is particularly relevant in the context of NPIs, 
which can lead to social dilemmas depending on the individual assessments \cite{traulsen:PNAS:2023, saad-roy:PNAS:2023, glaubitz2:arxiv:2023}.

Here, we examine the influence of decision-making on epidemic dynamics with variations in seasonal transmission rates, and vice versa. 
To investigate this, we extend a recent framework that couples individual adherence to an NPI with epidemic dynamics \cite{saad-roy:PNAS:2023} by introducing a seasonal transmission rate.
We associate adherence to an NPI to cooperation and not adhering to defection \cite{traulsen:PNAS:2023,glaubitz2:arxiv:2023,karlsson:SciRep:2020}. 
We then examine whether seasonality can introduce a tension between individual and population-level outcomes, i.e., a social dilemma, 
and whether individual behavior affects the impact of seasonality on epidemiological trajectories. 
Social dilemmas can arise depending on the infection levels in the population and the costs for adherence. 
With this approach, different seasons can be characterized by different social games -- even if we assume a constant number of infections, seasonality in disease transmission 
can imply that adherence to an NPI may be, e.g., a PD in spring and autumn, a SD game in winter and not a dilemma at all in summer. 
As we show below, coupling seasonal disease dynamics with behavioral dynamics can lead to complex und unexpected dynamics.

\section{Model}
 
\begin{figure*}[t]
    \centering
    \includegraphics[width=\linewidth]{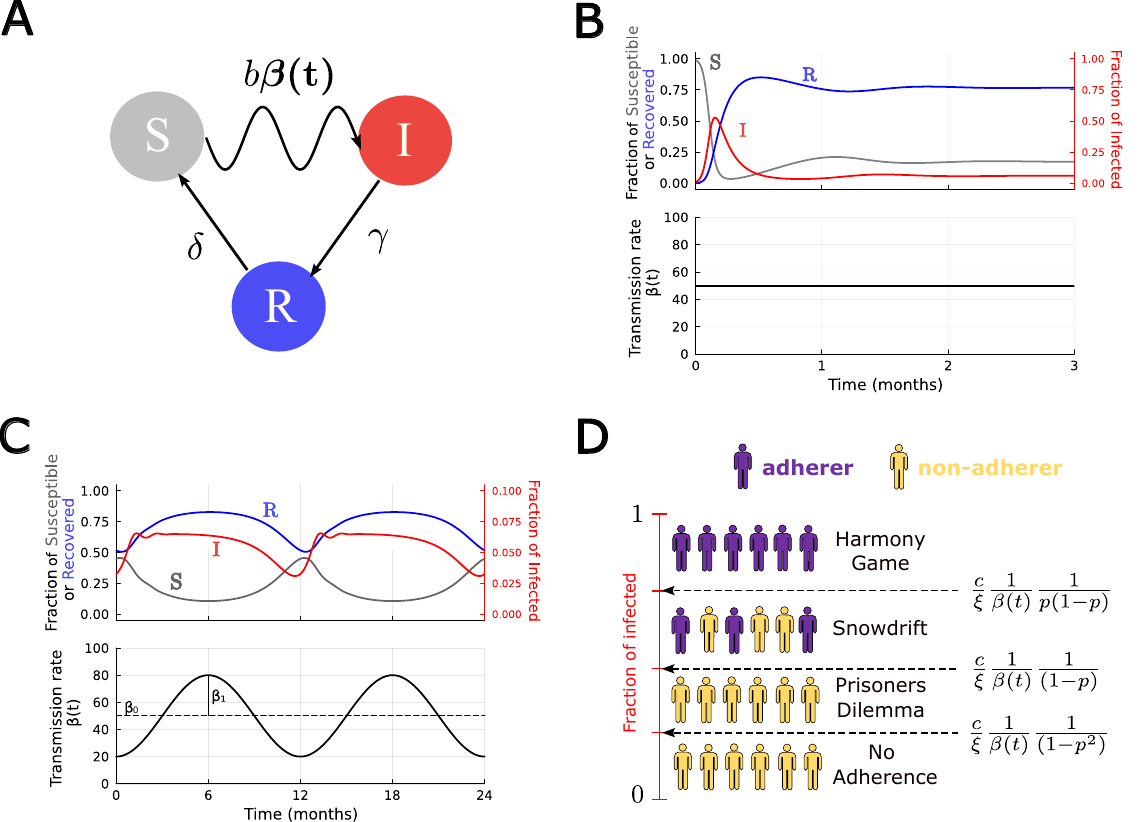}
    \caption{
    \sffamily
    {\bf Illustration of our model.} 
    \textbf{(A)} Schematic of the SIRS model, describing susceptible (S),
    infected (I), and recovered (R) and the transitions between them. 
    The seasonal transmission rate is $\beta(t)$, $b\leq 1$ is the average
    effect of adherence, $\gamma$ is the recovery rate and $\delta$
    is the rate at which recovered individuals become susceptible again (waning immunity).
    \textbf{(B)} SIRS model with constant transmission rate.
    We show an outbreak of the infection, transient oscillations and the dynamics approaching an endemic equilibrium for the 
    fraction of infected, susceptible, and recovered individuals as a function of time for a constant transmission rate 
    (parameters: $\beta=50, \gamma= 52/6, \delta=8/12$).
     \textbf{(C)} The SIRS model with seasonal transmission rate.
    We show the fraction of infected, susceptible, and recovered individuals as a function of time for a seasonal transmission rate after reaching the steady state,
    where $\beta_0$ is the baseline transmission rate, and $\beta_1$ the strength of oscillations. 
    Increasing the oscillation strength, the number of peaks in the infection change with a damping appearing for high enough values
    (parameters: $\beta_0=50, \beta_1= 30 , \gamma= 52/6, \delta=8/12$).
    \textbf{(D)} Behavioral dynamics. 
    We show a summary of all dilemma conditions depending on the fraction of infected individuals in the population.
    Non-adherers are depicted in yellow, and adherers in purple.
    Each game has a different level of adherence at the steady state,
    where no one adheres in the NA and PD games, some adhere in the SD game, and everyone adheres in the HG game.
    We show the thresholds for the number of infected individuals separating the games.
    }
    \label{model}
\end{figure*}

We consider a SIRS model, where a population is composed of
susceptible (S), infected (I) and recovered (R) individuals. The system of equations is given by 
\begin{subequations}
\begin{align}
    \diff{S}{t} &= - b \beta(t) S I + \delta R, \\
    \diff{I}{t} &=   b \beta(t) S I - \gamma I, \label{infected}\\
    \diff{R}{t} &=  \gamma I - \delta R,
\end{align}
\end{subequations}
where $\beta(t)$ is the transmission rate of the infection, 
$\gamma$ is the recovery rate and
$\delta$ is the rate of waning immunity.
Assuming a constant population size, we can set $S + I + R = 1$.
We define the seasonality of the transmission rate as
\begin{align}
     \beta(t) &= 
     \beta_0 - \beta_1 \cos \left(\tfrac{ 2\pi }{12} t \right) 
     \label{beta}
\end{align}
with a period of oscillations of twelve months (one year). 
We fix time such that the minimum $\beta_0 - \beta_1$ occurs in month 0. 
Here, $\beta_0$ is a baseline transmission rate and $\beta_1$ the strength of seasonality.
We illustrate the infection model in Fig.~\ref{model}A.
 
In addition to the infection dynamics in the population, we consider that individuals can adhere to an NPI. 
The average reduction in transmission due to adherence is captured by
\begin{align}
    b &= p x_A + (1-x_A).
\end{align}
We assume that $x_A$ is the probability of an individual to adhere to an NPI,
and $p$ is the effectiveness of adherence ($0<p<1$).
We define the payoffs for adherers (A) and non-adherers (N), given by the payoff matrix 
\begin{align}
    \bordermatrix{
                  & A & N \cr
                A & -p^2 \beta(t) \xi I - c & -p \beta(t) \xi I  - c  \cr
                N & -p \beta(t) \xi I &  - \beta(t) \xi I  \cr} .
                \label{adherence_matrix}
\end{align}
The individual payoffs depend on the fraction of infected individuals $I$, 
the risk perception $\xi$, the effectiveness of adherence $p$, the cost $c$ of following the NPI, and the transmission rate $\beta(t)$.
If no one adheres to the NPI, the transmission rate is $\beta(t)$.
If an adherer interacts with a non-adherer, the transmission rate is reduced by a factor of $p$. 
If both individuals adhere, the transmission rate is further reduced by a factor of $p^2$.

Thus, the average payoff of non-adherers is given by
\begin{align}
    \Pi_{N} &= \left[ -p \beta(t) \xi I \right] x_A + \left[ - \beta(t) \xi I \right] (1 - x_A)  ,
\end{align}
whereas the payoff of adherers is
\begin{align}
    \Pi_{A}  
        &= p \Pi_N -c \,.
\end{align}

We assume that the change in the probability to adhere follows a replicator equation with a rate of spontaneous switching $\mu$, leading to the equation
\begin{align}
      \frac{1}{\tau_A} \diff{x_A}{t} &=  x_A (1 - x_A ) (\Pi_{A} - \Pi_{N} )\\
     &+ \mu (1-x_A) - \mu x_A , \nonumber 
\end{align}
where $\tau_A$ is the time scale of the dynamics.
We introduce a spontaneous switching rate $\mu$ to prevent the population from reaching
absorbing states where behavior becomes fixed.

Based on the entries of a general payoff matrix
\begin{align}
    \bordermatrix{
                  & A & N \cr
                A & a_{AA} & a_{AN}  \cr
                N & a_{NA} &  a_{NN}  \cr},
\end{align}
different social games can be defined. 
Using the values in the adherence payoff matrix (Eq.~\ref{adherence_matrix}), we can determine the thresholds separating these games.
These boundaries depend on the effectiveness of adherence $p$, the risk perception $\xi$, the cost $c$ of adhering to the NPI, and the seasonal transmission $\beta(t)$.
We find the following thresholds:

\begin{itemize}
\item[(i)] If $a_{NA}>a_{AA}$, $a_{NN}>a_{AN}$, and $a_{NN}>a_{AA}$, non-adherence is always preferable. We will call this 
scenario as the No Adherence game (NA). This case arises when the fraction of infected individuals is
sufficiently small, 
\begin{align}
    I < \frac{c}{\xi} \frac{1}{\beta(t)} \frac{1}{(1-p^2)}.
\end{align}

\item[(ii)] If $a_{NA}>a_{AA}$, $a_{NN}>a_{AN}$, but $a_{NN}<a_{AA}$, non-adherence is preferable, but mutual adherence is better than mutual 
non-adherence. 
This defines a PD game, resulting in the condition
\begin{align}
    \frac{c}{\xi} \frac{1}{\beta(t)} \frac{1}{(1-p^2)} < I < \frac{c}{\xi} \frac{1}{\beta(t)} \frac{1}{(1-p)}.
\end{align}
In this case, any individual favors $N$ over $A$, even if the social optimum is that all individuals adhere. 

\item[(iii)] If $a_{NA}>a_{AA}$, $a_{NN}<a_{AN}$, and $a_{NN}<a_{AA}$, we have a SD game, where the population settles in a mixed configuration of N and A. 
The boundaries for this game are 
\begin{align}
    \frac{c}{\xi} \frac{1}{\beta(t)} \frac{1}{(1-p)} < I < \frac{c}{\xi} \frac{1}{\beta(t)} \frac{1}{p(1-p)}.
\end{align}
In this case, the social optimum is still that all individuals adhere, but individuals may still favor $N$ over $A$.

\item[(iv)] Finally, if 
$a_{NA}<a_{AA} $, $a_{NN}<a_{AN}$ and $a_{NN}<a_{AA}$, 
adherence is always favored.
This defines the Harmony Game (HG), which emerges for high fractions of infected individuals,  
\begin{align}
    I > \frac{c}{\xi} \frac{1}{\beta(t)} \frac{1}{p(1-p)}.
\end{align}
\end{itemize}
We summarise these thresholds in Fig.~\ref{model}D.
For $0<p<1$, the Stag-Hunt game cannot be reached -- this requires more sophisticated assumptions on the effect of an NPI \cite{traulsen:PNAS:2023}.
As the adherence dynamics follows the replicator equation, for $\mu \rightarrow 0$,
the population will converge to no adherence for the NA and PD game, 
partial adherence for the SD, 
and complete adherence in the HG \cite{nowak:book:2006}. 

For high transmission rates $\beta$ and high costs of adherence $c$,
the HG can only be reached for intermediate effectiveness $p$ \cite{saad-roy:PNAS:2023}.
Basically, if adherence is too effective, there is no need for 
everyone to adhere.
If adherence is too ineffective, it is not worth to do so.
For our numerical considerations, we will use $\gamma=52/6$, $\delta=8/12$, $c=52/6$
per month, $\xi=9$, and $p=0.5$.
This choice of parameter values allow us to encompass all possible social games dynamics,
and the disease parameter values are adopted from \cite{saad-roy:PNAS:2023}.

\section{Results}

\subsection{SIRS model with constant transmission}

First, we illustrate the SIRS model with a constant transmission rate and without
the introduction of adherence to an NPI (i.e., $b=1$).
A common way to assess the growth of an infection is by defining the effective reproductive number
$R_t$, 
$$R_t = \frac{\beta S}{\gamma}.$$
From Eq.~\eqref{infected}, we see that the fraction of infected individuals will grow
if $I (\beta S - \gamma)>0$.
Supposing that the infection is already present in the population ($I > 0$), this corresponds to $R_t > 1$. 
Initially, when most of the population is still susceptible ($S \approx 1$),
an outbreak occurs if $R_0 > \frac{\beta}{\gamma}$. 
In Fig.~\ref{model}B, we show the infection progression over time for this scenario. 
An outbreak occurs, creating a wave of infection in the initial steps, and
eventually establishing in the population.
In this model, an endemic equilibrium is possible due to the influx of susceptibles in the population, a consequence
of the waning immunity $\delta$. 
A similar dynamics occurs when considering birth and death processes \cite{saad-roy:PNAS:2023}. 

\subsection{SIRS model with seasonal transmission}

Next, we consider a seasonal transmission rate \cite{earn:science:2000}, according to Eq.~\eqref{beta}.
In Fig.~\ref{model}C, we show the oscillations in the fraction of susceptible, infected and recovered individuals after reaching a steady state.
These oscillations are driven by a seasonally oscillating transmission rate with a period of 12 months, starting at its minimum at the beginning of the year, and peaking in the middle.
For sufficiently strong seasonality, several peaks of infection occur early in the year, each with decreasing amplitude.
The period of these oscillations ($T$) is much faster than the annual cycle and arises from the dynamics of the infections diseases. 
It can be calculated by analyzing perturbations around the steady state of a system with constant $\beta$
(see Appendix \ref{sec:oscillations}).
For our seasonal transmission $\beta(t)$, we found numerically $T \approx 1.36$ months,
which is consistent with the period $T \approx 1.27$ months that can be calculated analytically from a constant transmission rate $\beta$.

\subsection{Seasonal SIRS model with NPI adherence}

\begin{figure}[tbp]
    \centering
    \includegraphics[width=\linewidth]{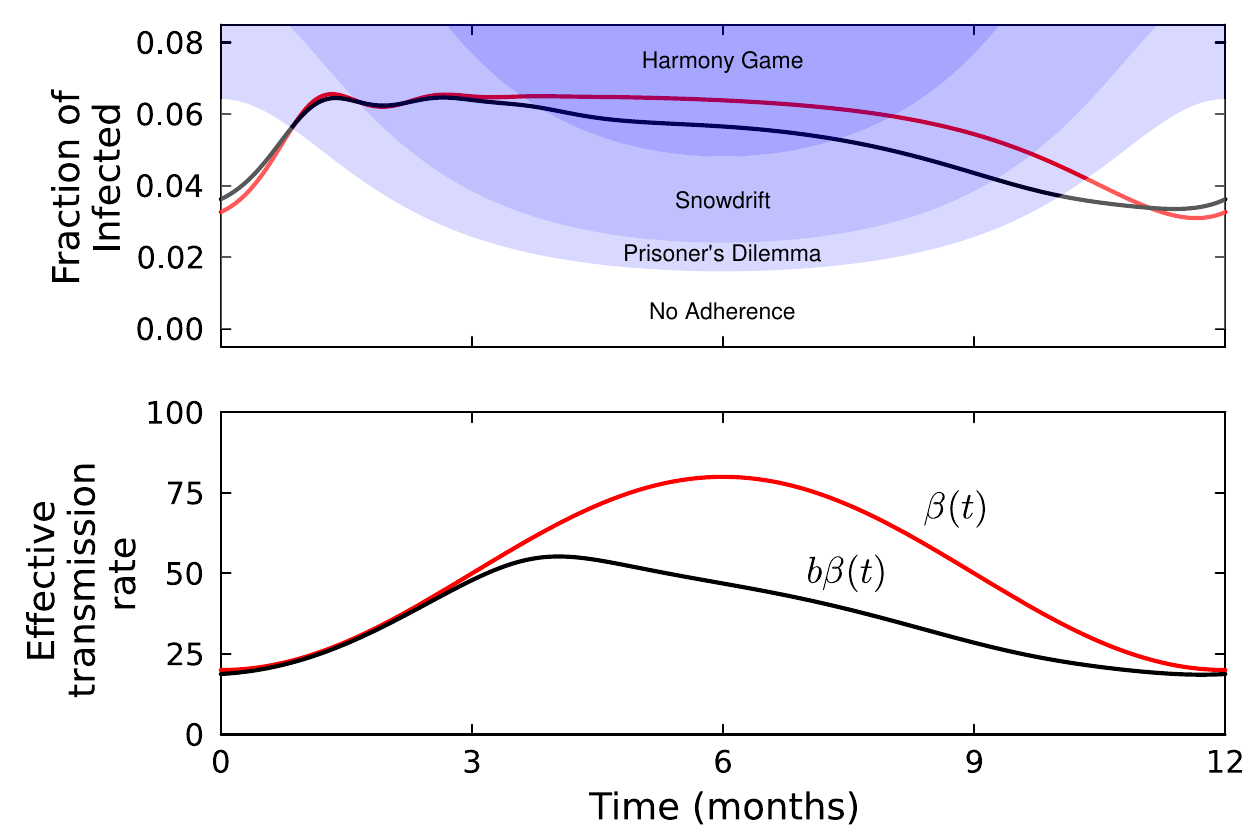}
    \caption{
    \sffamily
        {\bf Dynamic social games.}
        The fraction of infected individuals oscillates after reaching the steady state. 
        Red curves represent the SIRS model without adherence dynamics, while black curves represent the SIRS model with adherence dynamics. 
        The background colors represent the different social games
        defined by the thresholds in Fig.~\ref{model}D. 
        Note that at the beginning and at the end of the year, the model with adherence dynamics shows a higher fraction of infected due to an increased fraction of susceptible individuals. 
        The bottom panel displays the transmission rate $\beta(t)$
        and the effective transmission $b \beta(t)$ due to adherence (parameters: 
        $\beta_0=50$, $\beta_1 = 30$, $p=0.5$, $\tau_A = 0.25$, $\mu=10^{-7}$).
        }
        \label{games}
    \end{figure}

Finally, we explore the consequences of having adherers in the population. 
In Fig.~\ref{games}, we illustrate how the behavioral and epidemiological components of the model are coupled.
In the top panel we show the fraction of infected individuals over the period of one year after reaching the steady state. 
The bottom panel shows the effective transmission rate of the infection, that varies according to the fraction of adherence in the population.
At the beginning of the year, the infection does not spread due to the minimum in the transmission rate. 
Consequently, there is no need for individuals to adhere and pay the associated cost to do so.
Although adherence would decrease the fraction of infected individuals, 
both the individual and social optima are to not adhere.
Therefore, there is no social dilemma in this scenario (NA region).
As the transmission rate starts to increase, 
the fraction of infected individuals grows and the game turns into a PD.
Here, the social optimum would be to adhere, but this does not happen due to the temptation to not adhere. 
For even higher transmission rates, the population reaches the SD game.
Now, according to the replicator equation, adherence should start to increase, 
but there are not enough adherers in the population 
to decrease the transmission rate significantly.
Adherence only starts to grow substantially in the middle of the year (HG region), when the transmission rate is highest.
As a consequence, this increase in adherence reduces the transmission of the infection, lowering its abundance in the population.
Similar to the beginning of the year, there is no dilemma in this period.
After the peak in transmissibility, the fraction of infected individuals decreases, and the population will go back to the SD, PD, and NA, respectively.

\subsubsection*{Time scales}

In the previous section, we have discussed that partial adherence is expected in the SD region.
Despite that, we only observe a substantial growth in adherence in the HG region. 
This is due to the slow time scale of adherence, which delays the individuals responses to changes in the social game. 
One would expect that changing adherence faster can alter this picture compared to the SIRS dynamics. 
To investigate this, we explore different values of the time scale of adherence dynamics, $\tau_A$.
A larger value of $\tau_A$ indicates a faster change of adherence, whereas a smaller value of $\tau_A$ indicates a slower change of adherence.

\begin{figure}[t]
    \centering
    \includegraphics[width=\linewidth]{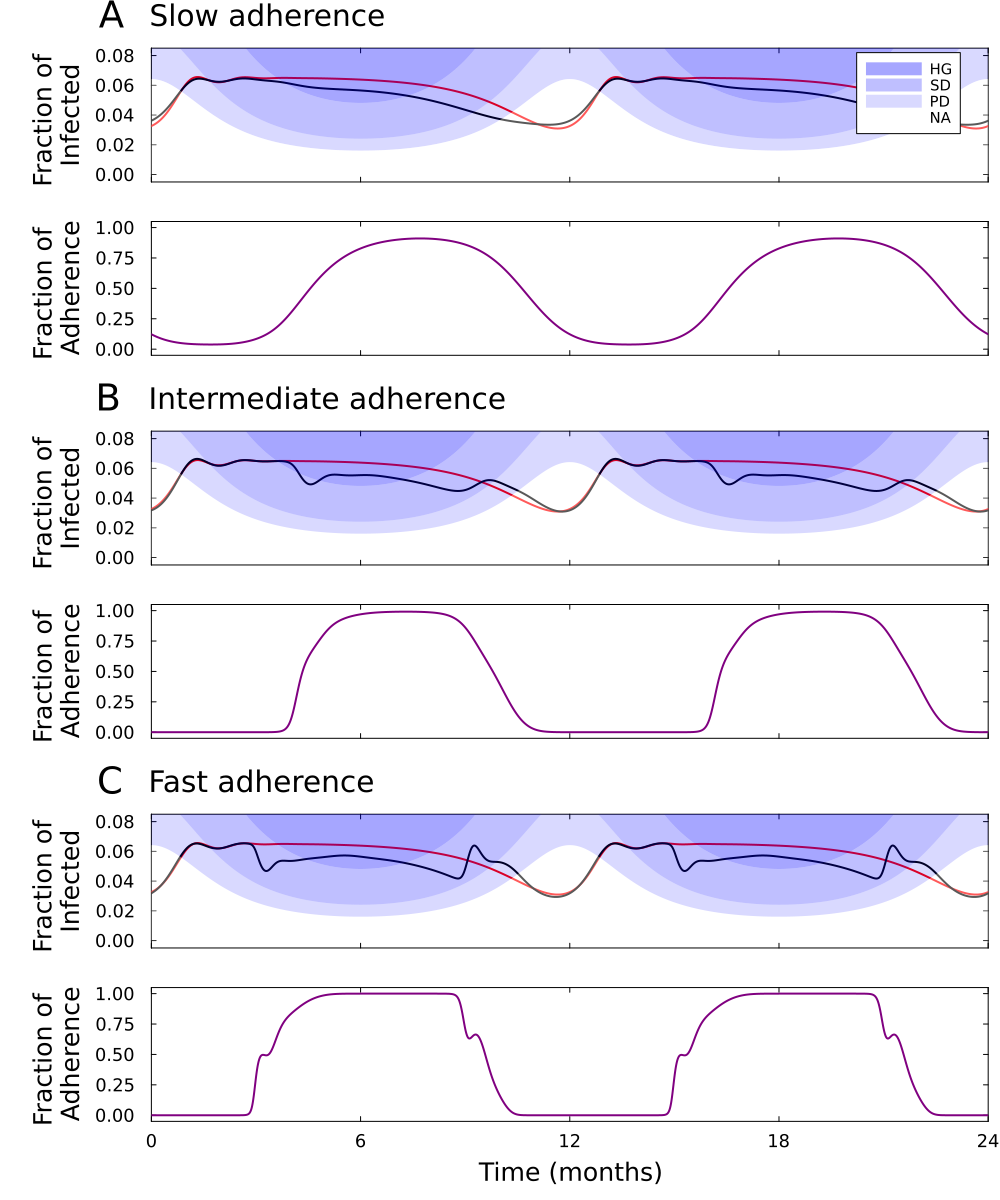}
    \caption{
    \sffamily
        {\bf Coupled model for behavior and seasonal epidemics}
        Fraction of infected and adherers over time after reaching the steady state for different time scales of adherence. 
        Red curves represent the SIRS without adherence, and black curves represent the SIRS with adherence. The background colors represent the different social games
        defined by the thresholds in Fig.~\ref{model}D.
        Panels (A-C) represent three time scales of adherence: slow, $\tau_A = 0.25$, intermediate, $\tau_A = 1$, and fast, $\tau_A = 4$.
        If the dynamics of adherence is slow, the peak in the fraction of infected can be avoided 
        when the transmission rate goes down towards the end of the year (panel A). 
        Similarly, by making adherence faster, the peak of infection increases (panel C)
        (The remaining parameters are $\beta_0=50$, $\beta_1 = 30$, $p=0.5$, $\mu=10^{-7}$). 
        }
        \label{time_scales}
    \end{figure}

Fig.~\ref{time_scales} shows the infection and adherence dynamics over time
after reaching a steady state oscillations, for different adherence time scales.
If adherence is slow (Fig.~\ref{time_scales}A), individuals reactions to changes in the social game are delayed.
Therefore, transitioning from the SD to PD, adherence decreases only slowly, 
leading to a benefit in situations that would result in no adherence otherwise (NA and PD).
This results in a less severe outbreak at the end of the year.

When we increase the time scale of adherence (Fig.~\ref{time_scales}B,C), individuals react faster when entering a social dilemma,
adhering earlier in the year (SD region).
On the other hand, when the population transition out of the dilemma, individuals will stop adhering
faster.
This quick exposure of susceptibles to the actual transmission rate 
will generate a higher peak of infection at the end of the year.
Theoretically, the best scenario to minimize the infection would be to have
a fast growth of adherence at the beginning of the infectious season, and a slow decline in adherence at the end.

\subsubsection*{Switching rate}

Finally, we explore how the switching rate $\mu$ affects the time dynamics of the infection.
Without switching, a population can get trapped in either complete adherence or no adherence.
By introducing switching, there will always be individuals adhering when they should not and vice-versa,
which can have a substantial effect on behavioral dynamics \cite{traulsen:PNAS:2009}.

\begin{figure}[t]
    \centering
    \includegraphics[width=\linewidth]{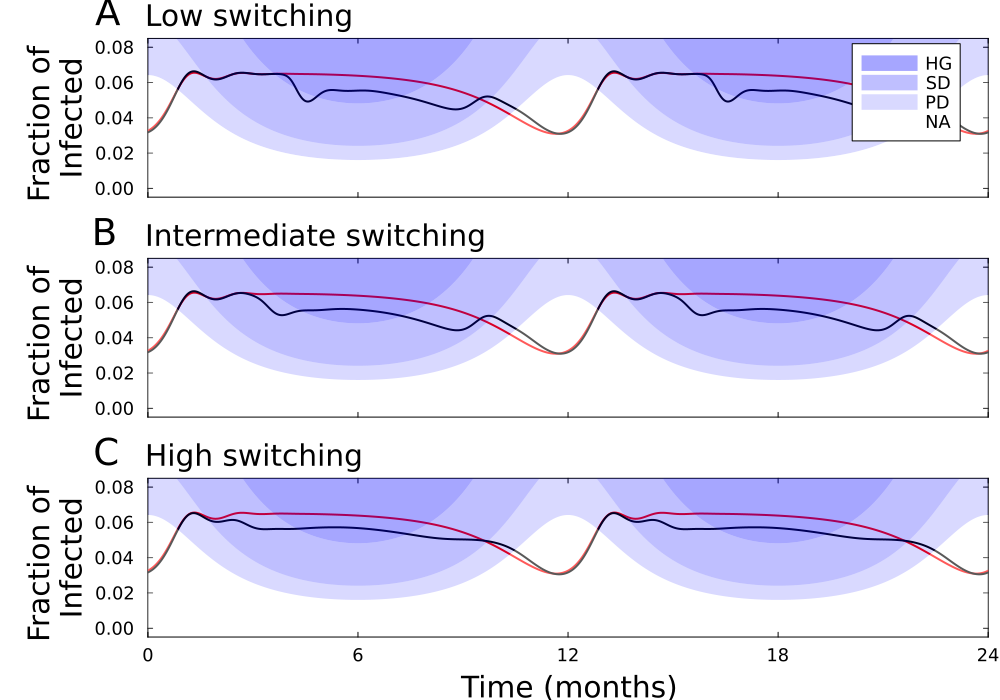}
    \caption{
    \sffamily
    {\bf Social dynamics with spontaneous switching.}
    Fraction of infected individuals over time after reaching the steady state for different switching rates $\mu$. 
    Red curves represent the SIRS without adherence, and black curves represent the SIRS with adherence and random switching between behaviours at rate $\mu$. 
    The background colors represent the different social games defined by the thresholds in Fig.~\ref{model}D.
    Panels (A-C) represent three different switching rates $\mu = 10^{-7}, 10^{-3}, 10^{-1}$, respectively.
    As the switching rate increases, it is possible to avoid the peak of infection at the end of the season, similarly to the slow adherence case.
    Furthermore, a higher switching rate also decreases the fraction of infected individuals earlier in the season
    compared to the SIRS model without adherence, similarly to the fast adherence case
    (parameters $\beta_0=50$, $\beta_1 = 30$, $p=0.5$, $\tau_A = 1$).
    }
    \label{mutations}
\end{figure}

Fig.~\ref{mutations} shows the fraction of infected individuals over time after reaching the steady state, for different values of $\mu$.
When the switching rate is low (Fig.~\ref{mutations}A), there are not many individuals reacting opposite to the social game. 
Thus, there are not many adherers at the beginning of the year, just as in the intermediate adherence case (Fig.~\ref{time_scales}B).
As the switching rate increases (Fig.~\ref{mutations}B,C), adherence starts to grow earlier in the year, 
a similar situation to the fast adherence case (Fig.~\ref{time_scales}C). 
At the same time, the peak of infection at the end of the year is avoided, resembling the slow adherence case (Fig.~\ref{time_scales}A).

By considering a spontaneous switching rate, partial adherence will be reached independently of the social game. 
Therefore, there are fewer infected individuals even in the NA and PD regions.
Also, since adherence never completely vanishes, the outbreak at the end of the year can be avoided.
On the other hand, complete adherence cannot be reached anymore, which is the most
beneficial scenario for decreasing transmission in the population.

\section{Discussion}

Epidemiological models allow us to understand and predict the spread of infectious diseases. 
At the same time, decision-making plays a crucial role in human interactions.
By coupling both subjects, we study a more complex situation, where the infection dynamics is affected by individual's choices to adhere to interventions that reduce transmission, and these choices are, in turn, based on the infection levels in the population.

We follow the usual convention of modeling a population composed of susceptible, infected and recovered individuals in the context of an infectious disease.
At the same time, individuals may adhere to an NPI with a certain probability, which decreases the transmission rate of the infection, but pay a cost to adhere.
The social optimum, which takes this cost into account, does not necessarily translate to the best epidemiological outcome.

Depending on the state of the infection in the population, adherence to an intervention can be a social dilemma.
By introducing a seasonal transmission rate, we have shown that different periods of the year are characterised by different social games and dilemmas.
If the infection is sufficiently severe, individuals will behave accordingly and will adhere to the intervention, thus decreasing infection levels.
However, if the infection is too mild, it is not worth to pay the cost of adherence.
In general, we observe that adherence to an NPI can decrease the infection levels in the population.
But on the other hand, increased adherence can also lead to more susceptible individuals in a population, leading to additional surges of the epidemic when adherence reduces.

More generally, periodically driven social games are still not well explored in the field of evolutionary game theory, with few works on the topic \cite{weitz:PNAS:2016}. 
Our work explores how such seasonal social dilemmas can arise from coupling decision-making and seasonal forcing of an infectious disease \cite{saad-roy:PNAS:2023,earn:science:2000,martinez:PLosPath:2018}, illustrating that such situations could naturally occur. 

Introducing mechanisms for the maintenance of cooperation in situations where the dilemma can be temporally reduced or even vanish will be an interesting area to be explored by evolutionary game theorists. 
Furthermore, more refined epidemiological models will benefit from taking into account how individuals react to the spread of an infectious disease \cite{bergstrom:PNAS:2023}. 
Ultimately, this will lead to a better understanding of human behavior in the context of changing risk.

\section*{Acknowledgments}
L.S.F. thanks the Brazilian funding agency 
CAPES (Coordenação de Aperfeiçoamento de Pessoal de Nível Superior) for his Ph.D. scholarship and the visitor program of the MPI for Evolutionary Biology for funding his research visit. 
A.A.-L. and A.T. acknowledge funding by the Collaborative Research Centre 1182: Origins and Functions of Metaorganisms.
C.M.S.-R. gratefully acknowledges funding from the Miller Institute of Basic Research in Science of the University of California, Berkeley via a Miller Research Fellowship.

\appendix
\section{Analysis of epidemiological oscillations \label{sec:oscillations}}

Here we analyze the oscillations in the SIRS model without adherence.
We consider the system of equations
\begin{align}
    \diff{S}{t} &= -\beta S I + \delta (1-S-I) \equiv f, \\
    \diff{I}{t} &=   \beta S I - \gamma I \equiv g,
\end{align}
with constant $\beta$ and $R=1-S-I$.
The endemic equilibrium ($S^*,I^*$) is given by 
$(S^*,I^*)=(\frac{\gamma}{\beta}, \frac{\delta}{\beta}\frac{\beta-\gamma}{\gamma+\delta})$.
Approximating the dynamics at the equilibrium we can write
\begin{align}
\diff{S}{t} &\approx (S - S^*) \frac{ \partial f}{\partial S} + (I - I^*) \frac{\partial f}{\partial I} \\
 \diff{I}{t} &\approx  (S - S^*) \frac{\partial g}{\partial S} + (I - I^*) \frac{\partial g}{\partial I}
\end{align}
Redefining the variables such that $\varepsilon_S = (S - S^*)$ and  $ \varepsilon_I = (I - I^*)$
we arrive at the system of equations
\begin{align}
    \diff{}{t} 
    \begin{bmatrix}
        \varepsilon_S \\  \varepsilon_I
    \end{bmatrix}
     &= J   \begin{bmatrix}
                \varepsilon_S \\  \varepsilon_I
            \end{bmatrix},
\end{align}
where $J$ is the Jacobian, given by
\begin{align*}
    J = \bordermatrix{
                  &  &  \cr
                 & \frac{\partial f}{\partial S} & \frac{\partial f}{\partial I}  \cr
                 & \frac{\partial g}{\partial S} & \frac{\partial g}{\partial I}  \cr}
        = \bordermatrix{
            &  &  \cr
           & -\beta I - \delta & -\beta S - \delta  \cr
           & \beta I  & \beta S - \gamma  \cr}.
\end{align*}
Calculating the eigenvalues of $J$ in the equilibrium ($S^*,I^*$) we find 
$\lambda_{\pm} = \lambda_r \pm i \lambda_i$, where
\begin{align}
    \lambda_r &=- \tfrac{1}{2} \tfrac{\beta  + \delta}{(\gamma+\delta)}\delta, \\
    \lambda_i & = 
 \tfrac{1}{2} \sqrt{\tfrac{-\delta  \left(\beta ^2 \delta -2 \beta  \left(2
   \gamma ^2+4 \gamma  \delta +\delta ^2\right)+4 \gamma ^3+8 \gamma ^2
   \delta +4 \gamma  \delta ^2+\delta ^3\right)}{(\gamma +\delta )^2}}
  \nonumber
   \\ 
   & 
    \approx \sqrt{\delta(\beta-\gamma)}
    \nonumber
   \\
&    \approx \sqrt{\beta \delta(1-S^*)}.
      \label{lambdaiApprox}
\end{align}
The real part is negative, such that the fixed point is stable. 
We approximated the imaginary part for $\delta \ll 1$ to have a more intuitive expression.
Since the fraction of susceptible is smaller than one, the imaginary part is nonzero,
such that the system reaches the fixed point in dampened oscillations.
The solution for the perturbation of the infected fraction of individuals around the equilibrium has the real part
\begin{align}
    \varepsilon_I (t) &\sim e^{(\lambda_r+ i \lambda_i) t} = e^{\lambda_r t} \cos(\lambda_i t) .
\end{align} 
The period $T$ of an oscillation of a cosine function is $2\pi$ divided by its argument,
\begin{align}
   T &= \frac{2 \pi}{\lambda_i} \approx 1.27
\end{align} 
using $\beta=50$, $\gamma=52/6$ and $\delta=8/12$, where time is measured in months.
Using the approximation for $\delta \ll 1$ Eq.~\eqref{lambdaiApprox}, we would instead obtain $T\approx 1.19$ months.

For the numerical solution of the differential equation with seasonally oscillating $\beta$, cf.~Eq.~\eqref{beta} we find from the distance between the two first peaks $T \approx 1.36$ months.

\end{document}